\documentclass{article}
\normalmarginpar
\twocolumn
\makeatletter
\gdef\@proofbox{\relax}
\long\def\proofbox#1{\gdef\@proofbox{#1}}
\proofbox{\small{\sl PhysComp96}\\\ifx\UNDEF\@fullpaper
Extended abstract\else Full paper\fi\\Draft, \@date}
\gdef\fullpaper{\gdef\@fullpaper{}}
\def\affil#1{\\{\small#1\par}}
\gdef\@author{John Doe1\affil{No-Name University, Shipping Dept.}}
\long\def\author#1{\gdef\@author{#1}}
\gdef\@abstract{}
\long\def\abstract#1{\gdef\@abstract{#1}}
\def\@maketitle{\newpage\leavevmode
  \begin{minipage}[t]{0.30\textwidth}
    \hrule height0pt
    \raggedright
    \mbox{}\par
    \@proofbox
  \end{minipage}\relax
  \begin{minipage}[t]{0.70\textwidth}
    \hrule height0pt
    \raggedleft
    \LARGE\@title\par
    \vskip4pt
    \large\@author
  \end{minipage}
  \vskip8pt
  \ifx\@abstract\@empty\else{\vskip.5em\leftskip1.5in\parskip4pt%
\small\@abstract\par\vskip.5em}\fi
  \rule{\textwidth}{0.4pt}
  \vskip16pt}

\let\df\bf
\DeclareRobustCommand\em
        {\@nomath\em \ifdim \fontdimen\@ne\font >\z@
                       \upshape \else \slshape \fi}

\def\@begintheorem#1#2{\sl \trivlist \item[\hskip \labelsep{\bf #1\ #2}]}
\def\@opargbegintheorem#1#2#3{\sl \trivlist
     \item[\hskip \labelsep{\bf #1\ #2\ (#3)}]}
\newcommand{\sect}[1]{\S\ref{sect.#1}}

\newcommand{\eq}[1]{(\ref{eq.#1})}

\newcommand{\sectlabel}[1]{\label{sect.#1}}
\newcommand{\eqlabel}[1]{\label{eq.#1}}

\def\@arabic#1{\number #1}
\makeatother

\fullpaper
\font\tenmsb=msbm10 scaled\magstep1
\font\sevenmsb=msbm7 scaled\magstep1
\font\fivemsb=msbm5 scaled\magstep1
\newfam\msbfam
\textfont\msbfam=\tenmsb
\scriptfont\msbfam=\sevenmsb
\scriptscriptfont\msbfam=\fivemsb
\def\Bbb#1{{\fam\msbfam\relax#1}}
\title{Tight bounds on\\quantum searching}
\author{Michel Boyer
\affil{Universit\'e de Montr\'eal\thanks{D\'epartement~IRO,
C.P.~6128, succursale centre--ville, Montr\'eal, \mbox{Canada H3C 3J7}.
\{boyer,brassard,tappa\}@iro.umontreal.ca}}
\par
Gilles Brassard, {\sc frsc}\thanks{Supported in part by
{\sc Nserc} and {\sc Fcar}}
\affil{Universit\'e de Montr\'eal}
\par
Peter H{\o}yer\thanks{Supported in part by the {\sc esprit}
Long Term Research Programme of the EU under project
number 20244 ({\sc alcom-it}).}
\affil{Odense University\thanks{Department of Mathematics and Computer Science,
Odense University, Campusvej~55, DK--5230 Odense M, Denmark.
u2pi@imada.ou.dk}}
\par
Alain Tapp\thanks{Supported in part by {\sc Nserc}}
\affil{Universit\'e de Montr\'eal}
}
\date{10 May 1996}
\abstract{We provide a tight analysis of Grover's recent algorithm
for quantum database searching.  We~give a simple
closed-form formula for the probability of success after any given number of
iterations of the algorithm.  This allows us to determine the number of
iterations necessary to achieve almost certainty of finding the answer.
Furthermore, we~analyse the behaviour of the algorithm when the element to be
found appears more than once in the table and we provide a new algorithm
to find such an element even when the number of solutions is not known
ahead of time.  Using techniques from Shor's quantum factoring algorithm
in addition to Grover's approach, we~introduce a new technique for approximate
quantum counting, which allows to estimate the number of solutions.
Finally we provide a lower
bound on the efficiency of any possible quantum database searching algorithm
and we show that Grover's algorithm nearly comes within a factor 2 of being
optimal in terms of the number of probes required in the table.}

\newcommand{\jqa}{\,.\,.\,}
\newcommand{\ket}[1]{\mbox{$| #1 \rangle$}}
\newtheorem{theorem}{Theorem}
\newtheorem{lemma}[theorem]{Lemma}
\newtheorem{proposition}[theorem]{Proposition}
\newcommand{\proof}{\noindent {\it Proof.}\ \ }
\newcommand{\qed}%
{\mbox{}\ \hspace*{\fill}\rule{1ex}{1.618ex}\vspace{\smallskipamount}}

\begin{document}

\maketitle

\section{Introduction}\sectlabel{intro}

Assume you have a large table $T[0 \jqa N-1]$ in which you would like to
find some element~$x$.  More precisely, you wish to find an integer $i$
such that \mbox{$0 \le i < N$} and \mbox{$T[i]=x$}, provided such an
$i$ exists.
This problem can obviously be solved in a time in $O(\log N)$ if the
table is sorted, but no classical algorithm (deterministic or
probabilistic) can succeed in the general case---when the elements
of $T$ are in an arbitrary order---with probability better than~$1/2$, say,
without probing more than half the entries of~$T$.
Grover~\cite{grover} has recently discovered an algorithm for the {\em quantum}
computer that can solve this problem in expected time in~$O(\sqrt{N\,}\,)$.
He~also remarked that a result in~\cite{BBBV} implies that
his algorithm is optimal, up to a multiplicative constant, among
all possible quantum algorithms.

In this paper we provide a tight analysis of Grover's algorithm.
In~particular we give a simple closed-form formula for the probability
of success after any given number of iterations.
This allows us to determine the number of iterations necessary
to achieve almost certainty of finding the answer, as well as
an upper bound on the probability of failure.
More significantly, we~analyse the
behaviour of the algorithm when the element to be found appears more
than once in the table.  An~algorithm follows immediately to solve
the problem in a time in $O(\sqrt{N/t\,}\,)$ when it is known that
there are exactly $t$ solutions.  We~also provide an algorithm
capable of solving the problem in a time in $O(\sqrt{N/t\,}\,)$
even if the number $t$ of solutions is not known in advance.
Bringing ideas from Shor's quantum factorization algorithm~\cite{shor} into
Grover's algorithm, we~sketch a new quantum algorithm capable of
approximately {\em counting} the number of solutions.
We~also generalize Grover's algorithm in the case $N$ is not a power of~2.
Finally, we refine the argument of~\cite{BBBV} to show that
Grover's algorithm could not be improved to require much less
than half the number of table lookups that it currently makes
when a 50\% probability of success is desired.

\section{Finding a unique solution}\sectlabel{unique}

Assume for now
that there is a unique $i_0$ such that
\mbox{$T[i_0]=x$}. For~any real numbers $k$ and $\ell$ such that
\mbox{$k^2+(N-1)\ell^2=1$}, define the state of a quantum register
\[  \ket{\Psi(k,\ell)} = 
    k \ket{i_0} + \sum_{i\neq i_0} \ell \ket{i} \]
where the sum is over all \mbox{$i \neq i_0$} such that \mbox{$0 \le i < N$}.
(We~shall never need complex amplitudes in this paper,
except in \sect{lower_bound}.)

The~heart of Grover's algorithm is a process,
henceforth called an {\em iteration}, that efficiently transforms
\ket{\Psi(k,\ell)} into
\ket{\Psi(\frac{N-2}{N} k + \frac{2 (N-1)}{N} \ell,
\frac{N-2}{N} \ell - \frac{2}{N} k)}.
Although we review the iteration process in~\sect{implementation}---where
we call it~$G$---we~refer
the reader to Grover's original article~\cite{grover}
for a more complete description and the proof that it performs as required.
Grover's algorithm begins by creating an equal superposition
\[ \ket{\Psi_0} = \ket{\Psi(1/\sqrt{N},1/\sqrt{N}\,)} =
\sum_{i=0}^{N-1} \frac{1}{\sqrt{N}} \, \ket{i} \]
of all possible values for~$i$, \mbox{$0 \le i < N$}.
Then some number $m$ of iterations are performed.
It~is clear from the above discussion that
the effect of the \mbox{$j$--th} iteration is to produce state
\mbox{\ket{\Psi_j}=\ket{\Psi(k_j,\ell_j)}} where \mbox{$k_0=l_0=1/\sqrt{N}$}
and
\begin{equation}\eqlabel{master}
\left.
\begin{array}{lll}
k_{j+1} &=& \frac{N-2}{N} k_j + \frac{2 (N-1)}{N} \ell_j \\[3mm]
\ell_{j+1} &=& \frac{N-2}{N} \ell_j - \frac{2}{N} k_j 
\end{array}
\right\}
\end{equation}
Finally, state \ket{\Psi_m} is observed, yielding some value~$i$.
The~algorithm {\em succeeds} if and only if \mbox{$T[i]=x$}.

In~his paper, Grover proves that there exists a number $m$ less than
$\sqrt{2N}$ such that the probability of success after $m$ iterations
is at least~$1/2$.  This is correct, but one must be careful in using
his algorithm because the probability of success does not increase
monotonically with the number of iterations.  By~the time you have
performed $\sqrt{2N}$ iterations, the probability of success has
dropped down to less than $9.5\%$ and it becomes vanishingly
small after about 11\% more iterations before it picks up again.
This shows that
it is not sufficient to know the existence of $m$ in order to
apply the algorithm in practice: its explicit value is needed.

The key to a tighter analysis of Grover's algorithm is an explicit
closed-form formula for $k_j$ and~$\ell_j$.  This can be obtained
by standard techniques---and a little sweat---from recurrence~\eq{master}.
Let angle $\theta$
be defined so that \mbox{$\sin^2 \theta = 1/N$}.  It~is straightforward
to verify by mathematical induction that
\begin{equation}\eqlabel{simple}
\left.
\begin{array}{lll}
k_j &=& \sin((2j+1) \theta) \\[3mm]
\ell_j &=& \frac{1}{\sqrt{N-1\,}} \cos((2j+1) \theta)
\end{array}
\right\}
\end{equation}
It follows from equation~\eq{simple} that \mbox{$k_m=1$}
when \mbox{$(2m+1) \theta=\pi/2$}, which happens when
\mbox{$m=(\pi-2\theta)/4\theta$}.  Of~course, we must perform
an {\em integer} number of iterations but it will be shown in
the next section that the probability of failure is no more
than $1/N$ if we iterate \mbox{$\lfloor \pi/4 \theta \rfloor$} times.
This is very close to \mbox{$\frac{\pi}{4} \sqrt{N}$}
when $N$ is large because \mbox{$\theta \approx \sin \theta = 1/\sqrt{N}$}
when \mbox{$\theta$} is small.
It~is sufficient to perform half this number of iterations,
approximately \mbox{$\frac{\pi}{8} \sqrt{N}$}, if we are content
with a 50\% probability of success, as Grover considered in his
original paper~\cite{grover}.
However, if we work twice as hard as we would
need to succeed with almost certainty, that is we apply
approximately $\frac{\pi}{2} \sqrt{N}$ iterations of Grover's algorithm,
we achieve a negligible probability of {\em success}!

\section{The case of multiple solutions}\sectlabel{known}

Let us now consider the case when there are $t$ solutions to the problem,
that is there are $t$ different values of $i$ such that \mbox{$T[i]=x$}.
We~are interested in finding an arbitrary solution.  Grover briefly
considers this setting~\cite{grover}, but he provides no details
concerning the efficiency of his method. 

We~assume in this section 
that the value of $t$ is known.  Let~\mbox{$A=\{i \,|\, T[i]=x \}$}
and \mbox{$B=\{i \,|\, T[i]\neq x \}$}.
For~any real numbers $k$ and $\ell$ such
that \mbox{$t k^2+(N-t)\ell^2=1$}, redefine
\[\ket{\Psi(k,\ell)} =
  \sum_{i\in A} k \ket{i} + \sum_{i\in B} \ell \ket{i} \, .\]
A straightforward analysis of Grover's algorithm shows
that one iteration transforms \ket{\Psi(k,\ell)} into
\[ \ket{\Psi(\frac{N-2t}{N} k + \frac{2 (N-t)}{N} \ell,
\frac{N-2t}{N} \ell - \frac{2t}{N} k)} \, . \]
This gives rise to a recurrence similar to~\eq{master},
whose solution is that the state \ket{\Psi(k_j,\ell_j)}
after $j$ iterations is given~by
\begin{equation}\eqlabel{sosimple}
\left.
\begin{array}{lll}
k_j &=& \frac{1}{\sqrt{t\,}}\sin((2j+1) \theta) \\[3mm]
\ell_j &=& \frac{1}{\sqrt{N-t\,}} \cos((2j+1) \theta)
\end{array}
\right\}
\end{equation}
where the angle $\theta$ is chosen so that \mbox{$\sin^2 \theta = t/N$}.

The~probability of obtaining a solution
is maximized when $\ell_m$ is as close to 0 as possible.
We~would have \mbox{$\ell_{\tilde{m}}=0$} when
\mbox{$\tilde{m}=(\pi-2\theta)/4\theta$} if that were an integer.  
Let~\mbox{$m=\lfloor \pi/4 \theta \rfloor$}.  Note that
\mbox{$|m-\tilde{m}|\le 1/2$}.  
It~follows that \mbox{$|(2m+1)\theta - (2\tilde{m}+1)\theta| \le \theta$}.
But~\mbox{$(2\tilde{m}+1)\theta=\pi/2$} by definition of $\tilde{m}$.
Therefore \mbox{$|\cos((2m+1) \theta)| \le |\sin \theta|$}.
We~conclude that the probability of failure after exactly $m$ iterations is
\[ (N-t) \ell_m^2 = \cos^2((2m+1) \theta) \le \sin^2 \theta = t/N \, . \]
This is negligible when \mbox{$t \ll N$}.

Note that this algorithm runs in a time in $O(\sqrt{N/t\,}\,)$
since \mbox{$\theta \ge \sin \theta = \sqrt{t/N\,}$}
and therefore
\[m \le \frac{\pi}{4\theta} \le
      \frac{\pi}{4} \sqrt{\frac{N}{t}} \, . \]

A slight improvement is possible in terms of the expected time
if we stop short of $m$ iterations, observe the register, and
start all over again in case of failure.  The~expected number
of iterations before success with this strategy is
\mbox{$E(j)=j/t k_j^2$} if we stop after $j$ iterations since our
probability of success at that point is~$t k_j^2$.
Setting the derivative of $E(j)$ to $0$ tells us that the
optimal number of iterations is given by the $j$ so that
\mbox{$4 \theta j = \tan((2j+1)\theta)$}.

We~have not solved this equation exactly but it is
very close to \mbox{$z=\tan(z/2)$} with \mbox{$z=4 \theta j$}
when the optimal $j$ is large, which happens when \mbox{$t \ll N$}.
The~solution for $z$ is approximately $2.33112$.
It~follows that the optimal number of iterations
is close to $0.58278 \sqrt{N/t}$ when \mbox{$t \ll N$}
and the probability of success is close to
\mbox{$\sin^2 (z/2) \approx 0.84458$}.
Therefore, the expected number of iterations
before success if we restart the process in case of failure is
roughly \mbox{$(z/(4 \sin^2 (z/2))) \sqrt{N/t} \approx 0.69003 \sqrt{N/t}$},
which is about 88\% of \mbox{$\frac{\pi}{4} \sqrt{N/t}$},
the number of iterations
after which success is almost certain.
For~a numerical example, consider the case \mbox{$N=2^{20}$}
and \mbox{$t=1$}.  In~this case, we achieve almost certainty
of success after 804 iterations.  If, instead, we stop at
596 iterations, the probability of success is only 0.8442
but the expected number of iterations before success if
we restart the process in case of failure is
\mbox{$596/0.8442 \approx 706$}, which is indeed better
than~$804$.

\subsection{The case {\protect\boldmath $t=N$}/4}

An interesting special case occurs when \mbox{$t=N/4$}.
Of~course, even a classical computer can find a solution
efficiently in this case, with high probability,
but not quite as efficiently
as a quantum computer.  Here \mbox{$\sin^2 \theta = t/N = 1/4$}
and therefore \mbox{$\theta=\pi/6$}.  This implies that
\[ \ell_1 = \frac{1}{\sqrt{N-t\,}} \cos(3 \theta) = 0 \, . \]
In~other words, a solution is found {\em with certainty}
after a single iteration.  Because one iteration of Grover's
algorithm requires two table look-ups (including one for
uncomputation purposes---see~\sect{lower_bound}), this is twice as efficient
(in terms of table look-ups) than the expected performance
of the obvious classical probabilistic algorithm---and that's
best possible classically.  Furthermore, the quantum algorithm
becomes {\em exponentially} better than any possible
classical algorithm if we compare worst-case performances,
taking the worst possible coin flips in the case of a
probabilistic algorithm.  This is somewhat reminiscent
of the Deutsch--Jozsa algorithm~\cite{DJ}.
 
\section{Unknown number of solutions}\sectlabel{unknown}

A~much more interesting case occurs when the number of solutions
is not known ahead of time.  If~we decide to iterate
\mbox{$\frac{\pi}{4} \sqrt{N}$} times, which would give almost
certainty of finding a solution if there were only one,
the probability of success would be vanishingly small should
the number of solutions be in fact 4 times a small perfect square.
For~example we saw above that we
are almost certain to find a unique solution among
$2^{20}$ possibilities if we iterate 804 times.
The~same number of iterations would yield a solution
with probability less than one in a million should there
be 4 solutions!  In~order to find a solution efficiently
when their number is unknown, we need the following lemmas,
the first of which is proved by straightforward algebra.

\begin{lemma}
For any real numbers $\alpha$ and $\beta$, and any positive
integer~$m$,
\[ \sum_{j=0}^{m-1} \cos(\alpha+2 \beta j) = 
   \frac{\sin(m \beta) \cos(\alpha+(m-1)\beta)}{\sin \beta} \, . \]
In particular, when $\alpha=\beta$,
\[ \sum_{j=0}^{m-1} \cos((2j+1)\alpha) =
   \frac{\sin(2 m \alpha)}{2 \sin \alpha} \, . \]
\end{lemma}

\begin{lemma}\label{lemma2}
Let $t$ be the (unknown) number of solutions and let
$\theta$ be such that \mbox{$\sin^2 \theta = t/N$}.
Let $m$ be an arbitrary positive integer.
Let $j$ be an integer chosen at random according to the
uniform distribution between 0 and \mbox{$m-1$}.
If~we observe the register after
applying $j$ iterations of Grover's algorithm
starting from the initial state 
\mbox{$\ket{\Psi_0}=\sum_i \frac{1}{\sqrt{N}} \ket{i}$},
the probability of obtaining a solution is exactly
\[ P_m ~=~ \frac{1}{2} - \frac{\sin(4 m \theta)}{4 m \sin(2 \theta)} \, . \]
In particular \mbox{$P_m \ge 1/4$} when \mbox{$m \ge 1/\sin(2\theta)$}.
\end{lemma}
\proof
The probability of success if we perform $j$ iterations of
Grover's algorithm is
\mbox{$t k_j^2 = \sin^2((2j+1) \theta)$}.
It~follows that the average success probability when
\mbox{$0 \le j < m$} is chosen randomly is
\begin{eqnarray*}
P_m &=& \sum_{j=0}^{m-1} \frac{1}{m} \sin^2((2j+1) \theta) \\*
 &=& \frac{1}{2m} \sum_{j=0}^{m-1} 1-\cos((2j+1) 2 \theta) \\*
 &=& \frac{1}{2} - \frac{\sin(4 m\theta)}{4 m \sin(2 \theta)} \, .
\end{eqnarray*}
If $m \ge 1/\sin(2 \theta)$ then
\[ \frac{\sin(4 m\theta)}{4 m \sin(2 \theta)} \le 
\frac{1}{4 m \sin(2 \theta)} \le \frac{1}{4} \, . \]
The conclusion follows.
\qed

We~are now ready to describe the algorithm for finding a solution
when the number $t$ of solutions is unknown.  For~simplicity we assume
at first that \mbox{$1 \le t \le 3N/4$}.
\begin{enumerate}
\item Initialize $m=1$ and set $\lambda=6/5$.\\
(Any value of $\lambda$ strictly between $1$ and $4/3$ would~do.)
\item choose $j$ uniformly at random among the nonnegative integers
smaller than~$m$.
\item Apply $j$ iterations of Grover's algorithm starting from
initial state \mbox{$\ket{\Psi_0}=\sum_i \frac{1}{\sqrt{N}} \ket{i}$}.
\item Observe the register: let~$i$ be the outcome.
\item If $T[i]=x$, the problem is solved: {\df exit}.
\item Otherwise, set $m$ to \mbox{$\min(\lambda m, \sqrt{N}\,)$}\\
and go back to step~2.
\end{enumerate}

\begin{theorem}
This algorithm finds a solution in expected time in
$O(\sqrt{N/t\,}\,)$.
\end{theorem}
\proof
Let $\theta$ be the angle so that \mbox{$\sin^2 \theta = t/N$}.
Let
\[ m_0 = 1/\sin(2\theta) = \frac{N}{2 \sqrt{(N-t)t}} < \sqrt{\frac{N}{t}} \]
(recall that we assumed $t \le 3N/4$).

We shall estimate the expected number of times that a Grover
iteration is performed: the total time needed is clearly
in the order of that number.  On~the \mbox{$s$--th} time round
the main loop, the value of $m$ is $\lambda^{s-1}$ and the
expected number of Grover iterations is less than half that value
since $j$ is chosen randomly so that \mbox{$0 \le j < m$}.
We~say that the algorithm reaches the {\em critical stage}
if it goes through the main loop more than
\mbox{$\lceil \,\log_{\lambda} m_0 \rceil$} times.
The~value of $m$ will exceed $m_0$ if
and when the algorithm reaches that stage.

The expected total number of Grover iterations needed to reach the
critical stage, if it is reached, is at most
\[ \frac{1}{2} \sum_{s=1}^{\lceil \,\log_{\lambda} m_0 \rceil} \lambda^{s-1}
 < \frac{1}{2} \frac{\lambda}{\lambda-1} m_0 = 3 m_0 \, . \]
Thus, if the algorithm succeeds before reaching the critical stage,
it does so in a time in $O(m_0)$, which is in $O(\sqrt{N/t}\,)$
as required.

If the critical stage is reached then every time round the main
loop from this point on will succeed with probability at least~$1/4$
by virtue of Lemma~\ref{lemma2} since \mbox{$m \ge 1/\sin(2\theta)$}.
It~follows that the expected number of Grover iterations needed to
succeed once the critical stage has been reached is
upper-bounded by
\[ \frac{1}{2} \sum_{u=0}^{\infty} \frac{3^u}{4^{u+1}}\,
         \lambda^{u+\lceil \,\log_{\lambda} m_0 \rceil} <
         \frac{\lambda}{8-6\lambda} m_0 = \frac{3}{2} m_0 \, . \]
The total expected number of Grover iterations, in case the critical stage is
reached, is therefore upper-bounded by $\frac{9}{2} m_0$
and thus the total expected time is in $O(\sqrt{N/t\,}\,)$
provided $0<t\le 3N/4$.
Note that \mbox{$\frac{9}{2} m_0 \approx \frac{9}{4}\sqrt{N/t}$}
when \mbox{$t \ll N$}, which is less than four times the expected
number of iterations that we would have needed had we known the
value of $t$ ahead of time.
The~case \mbox{$t>3N/4$} can be disposed of
in constant expected time by classical sampling.  The~case \mbox{$t=0$} is
handled by an appropriate time-out in the above algorithm, which allows to
claim in a time in $O(\sqrt{N}\,)$ that there are no solutions when this is
the case, with an arbitrarily small probability of failure when in fact there
is a solution.
\qed

\section{Quantum counting}\sectlabel{counting}

We are currently investigating the power of quantum computers in
approximately {\em counting} the number $t$ of solutions, rather than
merely finding~one.
For~this, we use techniques inspired by Shor's celebrated
quantum factorization algorithm~\cite{shor} and combine them
with Grover's algorithm.
Here we sketch the basic ideas, leaving the details---many of which still
have to be worked out---to~a further paper~\cite{future}.

Let $k_j$ and $\ell_j$ be as in equation~\eq{sosimple}
and recall that \mbox{$A=\{i \,|\, T[i]=x \}$}
and \mbox{$B=\{i \,|\, T[i]\neq x \}$}.
The key observation is that the value of~$\theta$,
and therefore that of~$t$, can be inferred directly from
the {\em period} of the function that sends $j$ onto~$k_j$.
This~period can be estimated from sampling in a discrete Fourier transform
of the function.  In~order to profit from the ability of quantum
computers to compute Fourier transforms, though, we must first create
a state in which the amplitude of \ket{j} is proportional to~$k_j$
for values of~$j$ ranging over several periods.

Let $P$ be a power of~2, arbitrary for the moment, and let
\mbox{$f= P \theta / \pi$} be the number
of periods of $k_j$ when $j$ spans the range from 0 to \mbox{$P-1$}.
(In~general $f$ need not be an integer.)
Create state
\[ \ket{\Psi_0} = \sum_{j=0}^{P-1} \sum_{i=0}^{N-1}
  \frac{1}{\sqrt{PN}} \ket{j,i} \, . \]
Then apply to \ket{\Psi_0} a transformation that sends
\ket{j}\ket{\Psi} to \mbox{$\ket{j}\,G^j\ket{\Psi}$},
where $G$ is the Grover iteration.  This takes a time proportional to~$P$,
resulting in the state
\[ \sum_{j=0}^{P-1} \left[ \frac{1}{\sqrt{P}} \ket{j} \left(
  \sum_{i \in A} k_j \ket{i} + \sum_{i \in B} l_j \ket{i} \right)
  \right] \, . \]
Now, observe the second part of the register.
Assume without loss of generality that some element from $A$ is obtained.
(There are no essential differences if instead an element from $B$
is obtained since $k_j$ and $\ell_j$ have exactly the same period.)
At~this point, the first part of the register has collapsed to state
\[ \sum_{j=0}^{P-1} k_j \ket{j} \]
up to renormalization.
If~we apply a quantum discrete Fourier transform to this state~\cite{shor}
({\em not}~what Grover calls the quantum Fourier transform in~\cite{grover}!),
and if $f$ is large enough,
the amplitude of all values of $j$ becomes vanishingly small, except
for values very close to $f$ or \mbox{$P-f$}.
\mbox{Finally}, we observe the register.  With high probability,
this yields an excellent approximation $\tilde{f}$ on~$f$, from
which we estimate
\[ \tilde{\theta} = \frac{\tilde{f} \pi}{P} ~~~\mbox{and}~~~
   \tilde{t} = N \sin^2 \tilde{\theta} \, . \]

To~evaluate the accuracy of~$\tilde{t}$, we~assume that
\mbox{$|f-\tilde{f}|<1$}, which happens with reasonable probability
provided $f$ is sufficiently large---see~\cite{future} for details.
It~follows that \mbox{$|\theta-\tilde{\theta}|<\pi/P$}
and therefore \mbox{$|\sin \theta-\sin \tilde{\theta}|<\pi/P$}
as well.  From \mbox{$\tilde{t}=N \sin^2 \tilde{\theta}$},
\mbox{$t=N \sin^2 \theta$} and \mbox{$\sin \theta = \sqrt{t/N}$},
we derive
\begin{equation}\eqlabel{error}
|t - \tilde{t}| < \frac{2\pi}{P} \sqrt{tN} + \frac{\pi^2}{P^2} N \,.
\end{equation}
Recall that the running time of the algorithm is proportional to~$P$.
This parameter allows us to balance the desired accuracy of the
approximation with the running time required to achieve~it.
Let~$c$ be a constant.
\begin{itemize}

\item[$\diamond$] If we take \mbox{$P=c\sqrt{N}$}, the error
in our estimate of $t$ is bounded by
\mbox{$\frac{2 \pi}{c} \sqrt{t} + \frac{\pi^2}{c^2}$}
provided \mbox{$|f-\tilde{f}|<1$}.
This is reminiscent of finding the answer up to a few standard
deviations.

\item[$\diamond$] If we are satisfied with keeping small
the {\em relative} error, we run the algorithm on successive powers
of~2 for~$P$ until $\tilde{f}$ becomes reasonably large.
This will happen when \mbox{$P=c\sqrt{N/t}$}.
After a total time proportional to $\sqrt{N/t}$,
this yields an estimate for~$t$ that is likely to be within
a factor \mbox{$(1+\pi/c)^2$} of the correct answer.

\item[$\diamond$] If we want the {\em absolute} error to be probably
bounded by a constant, we apply the algorithm once with \mbox{$P=c\sqrt{N}$}
in order to estimate~$t$.  Then, we run it again, but with
\mbox{$P=c\sqrt{\tilde{t}N}$}.  According to equation~\eq{error},
and pretending \mbox{$P=c\sqrt{tN}$} for simplicity, the resulting
error in our second estimate of $t$ is likely to be bounded
by \mbox{$\frac{2 \pi}{c} + \frac{\pi^2}{c^2 t}$}.
In~particular, we get the {\em exact} answer, provided
\mbox{$|f-\tilde{f}|<1$}, if we take \mbox{$c \ge 14$}
since \mbox{$\frac{2 \pi}{c} + \frac{\pi^2}{c^2 t} < 1/2$} in that case.
(Note that successive applications of Grover's algorithm in which
we strike out the solutions as they are found will also provide
an exact count with high probability in a time in $O(\sqrt{tN\,}\,)$,
but at an enormous cost in terms of additional memory---see~\cite{future}.)

\item[$\diamond$] Finally, we~have a variation on this technique that gives the
{\em exact} answer in a time in $O(\sqrt{N\,}\,)$ with a vanishingly small
probability of error provided the number of solutions is a small perfect
square.

\end{itemize}
We~defer the details to~\cite{future}.

\section{Implementation considerations}\sectlabel{implementation}

Grover's algorithm consists of a number of iterations followed by a 
measurement. In~his original article~\cite{grover} Grover shows that
the unitary transform $G$, defined below,
efficiently implements what we called an iteration in~\sect{unique}.

For every $A \subset \Bbb Z_N$, 
let $S_A$ be the conditional phase shift transform given by
\[S_A \ket{i} = \left\{ \begin{array}{rl}  - \ket{i} & \mbox{if } i \in A\\
\ket{i} & \mbox{otherwise.}
\end{array} \right.  \]
For every $i \in \Bbb Z_N$, denote
$S_{\{i\}}$ by $S_i$.
Let~$T$ be the Walsh-Hadamard transform
\[T \ket{j} = \frac{1}{\sqrt{N}} \sum_{i=0}^{N-1} (-1)^{i \cdot j} \ket{i},\]
where $i \cdot j$ denotes the bitwise dot product of the two 
strings~$i$ and~$j$.
Then the transform $G$ is given by
\[G = - T S_0 T S_{i_0}.\]

Grover considers only the case when $N$ is a power of~2
since the transform $T$ is well-defined only in this case. 
However,
the assumption on $N$ can be removed by observing that $G$ is 
just one of many transforms that efficiently implements an iteration.
Let $T'$ be any unitary transform satisfying
\begin{equation}\eqlabel{ttransform}
T' \ket{0} = \frac{1}{\sqrt{N}} \sum_{i=0}^{N-1} \ket{i}.
\end{equation}
Then one may easily verify that the transform 
$T' S_0 {T'}^{-1} S_{i_0}$
works just as well, and, more interestingly, that
\[ T' S_0 {T'}^{-1} S_A\]
implements the general iteration analysed in~\sect{known}.
Any transform $T'$ satisfying~\eq{ttransform} can thus be used in 
the algorithm. 

When $N$ is a power of~2, the Walsh-Hadamard transform
is indeed the simplest possible choice for $T'$. When $N$ is not a power
of two, the approximate Fourier transform given by 
Kitaev~\cite{kitaev} can be used.

\section{An improved lower bound}\sectlabel{lower_bound}
\newcommand{\mbf}[1]{\mbox{\bf #1}}
\def\inner#1#2{\langle#1, #2\rangle}

Drawing on general results from~\cite{BBBV}, Grover points out
that any algorithm for quantum database searching must take a time at least
proportional to $\sqrt{N}$ when there is a unique solution.  Here we
refine and generalize this result by giving an explicit lower bound on the
number of table lookups required by any quantum algorithm as a function of
the number of solutions.  This lower bound is only a few percent smaller
than the number of iterations required by Grover's algorithm when the
number of solutions is known in advance.  Unfortunately, each iteration of
Grover's algorithm requires {\em two} table lookups because $T[i]$ must
first be fetched (to~decide on potential phase shift) and then it must
be erased (to~allow interference to take place) by a process often referred
to as uncomputation.  Therefore, we merely prove that Grover's algorithm
is roughly within a factor~2 of being optimal in terms of the number
of table lookups.

We~rephrase the problem in terms of an oracle $O$ defined so that
$O(i)=1$ whenever $i$ is a solution.
All matrices and vectors in this section are finite and complex-valued.
Let the {\em inner product\/} $\inner{\mbf{a}}{\mbf{b}}$
of two vectors $\mbf{a}$ and $\mbf{b}$ be defined as 
$\sum_i a_i^\star b_i$, where $c^\star$ denotes the 
complex conjugate of~$c$.
The norm of $\mbf{a}$ is denoted $\|\mbf{a}\|$.
The absolute value of a complex number $c$ is denoted $|c|$.

We restate a basic fact on complex-valued vectors:
\begin{proposition}
\label{prop:1}
For all normalized vectors $\mbf{a}$ and $\mbf{b}$, 
and all complex scalars $\alpha$ and $\beta$,
\[  \|\alpha \mbf{a} - \beta \mbf{b}\|^2   \geq   
    |\alpha|^2 + |\beta|^2 - 2 |\alpha| |\beta|. \]
\end{proposition}

The following proposition is a consequence of Chebyshev's 
summation inequalities.
\begin{proposition}
\label{prop:2}
For all set of complex numbers, $\{x_i\}_{i=0}^{r-1}$,
\[ \left(\sum_{i=0}^{r-1} |x_i|\right)^2 
  \leq r \sum_{i=0}^{r-1} |x_i|^2. \]
\end{proposition}

\begin{lemma}
\label{lm:main:lower}
Let $S$ be any set of $N$ strings, 
and $\cal C$ be any configurationspace.
Let $\ket{\phi_0}$ be any superposition, and
$$\ket{\phi_r} = U_r \dots U_2 U_1 \ket{\phi_0}$$
any sequence of $r$ unitary transforms. 
Let $\{f_i\}_{i=0}^r$ be any set of partial functions from
$\cal C$ into $S$.
For any $y \in S$, let 
$$\ket{\phi'_r} = U'_r \dots U'_2 U'_1 \ket{\phi_0}$$
be any sequence of $r$ unitary transforms where for all \mbox{$i=1,\dots,r$},
$$U'_i \ket{c} = U_i \ket{c} 
   \quad \mbox{ if } \quad f_{i-1}(\ket{c}) \neq y.$$

Set $\ket{\phi'_0} = \ket{\phi_0}$, and for all $i=1,\dots,r$,
set $\ket{\phi_i}=U_i \ket{\phi_{i-1}} \mbox{ and }
\ket{\phi'_i}=U'_i \ket{\phi'_{i-1}}.$
For all $i=0,1,\dots,r$,
set $\ket{\phi_i}=\alpha_{i,y}\ket{\phi_{i,y}} + 
\alpha_{i,\overline{y}}\ket{\phi_{i,\overline{y}}}$, where $\ket{\phi_{i,y}}$
$(\ket{\phi_{i,\overline{y}}})$ is the normalized superposition of
configurations where $f_i$ (does not) equals $y$.
Denote $\ket{\phi'_i}$ similarly.

Then the following holds:
\begin{enumerate}
\item
$\| \ket{\phi'_r} - \ket{\phi_r} \| 
    \leq 2 \sum_{i=0}^{r-1} |\alpha_{i,y}| 
\quad \mbox{ for all } y \in S$
\item
$2 \{1 - |\alpha_{r,y}| - |\alpha'_{r,\overline{y}}|\}
    \leq \| \ket{\phi'_r} - \ket{\phi_r} \|^2 
\quad \mbox{ for all } y \in S$
\item
$N - \sqrt N - \sum_{y \in S} |\alpha'_{r,\overline{y}}| 
    \leq 2 r^2$
\end{enumerate}
\end{lemma}

\proof
We divide the proof into three parts.

\noindent
Proof of (1):
For all $y \in S$, and all $i=1,\dots,r$ we have
\[\begin{array}{l}
U'_i \ket{\phi_{i-1}} 
 = U'_i \left( \alpha_{i-1,y} \ket{\phi_{i-1,y}} + 
        \alpha_{i-1,\overline{y}} \ket{\phi_{i-1,\overline{y}}}\right) \\
= U'_i \left( \alpha_{i-1,y} \ket{\phi_{i-1,y}}\right)
    + U_i \left( \alpha_{i-1,\overline{y}} \ket{\phi_{i-1,\overline{y}}}\right)
\\
= U'_i \left( \alpha_{i-1,y} \ket{\phi_{i-1,y}}\right)
    - U_i \left( \alpha_{i-1,y} \ket{\phi_{i-1,y}}\right)  
    + U_i \ket{\phi_{i-1}} \\
= \ket{\phi_i} + (U'_i - U_i) 
      \left(\alpha_{i-1,y} \ket{\phi_{i-1,y}}\right).
\end{array} \]

Hence, by induction on $i$,
\[\begin{array}{l}
\ket{\phi'_i} = U'_i \dots U'_1 \ket{\phi_0} \\
= \ket{\phi_i} + \sum_{j=1}^i (U'_i \dots U'_{j+1})(U'_j - U_j) 
         \left(\alpha_{j-1,y} \ket{\phi_{j-1,y}}\right),
\end{array}\]
so,
\[\begin{array}{l}
\| \ket{\phi'_i} - \ket{\phi_i} \|  \\
= \| \sum_{j=1}^i  (U'_i \dots U'_{j+1})(U'_j - U_j) 
         \left(\alpha_{j-1,y} \ket{\phi_{j-1,y}}\right) \|  \\
\leq 2 \sum_{j=1}^i |\alpha_{j-1,y}|,
\end{array}\]
and (1) follows.

\noindent
Proof of (2):
The identity follows from:
\[\begin{array}{l}
\| \ket{\phi'_r} - \ket{\phi_r} \| \\
= \| (\alpha'_{r,y} \ket{\phi'_{r,y}}
          + \alpha'_{r,\overline{y}} \ket{\phi'_{r,\overline{y}}})\\
\phantom{= \| }  - (\alpha_{r,y} \ket{\phi_{r,y}}
          + \alpha_{r,\overline{y}} \ket{\phi_{r,\overline{y}}})\|\\
= \| (\alpha'_{r,y} \ket{\phi'_{r,y}}
    - \alpha_{r,y} \ket{\phi_{r,y}})  \\
\phantom{= \| }  + (\alpha'_{r,\overline{y}} \ket{\phi'_{r,\overline{y}}}
    - \alpha_{r,\overline{y}} \ket{\phi_{r,\overline{y}}})\|\\
= \big\{\|\alpha'_{r,y} \ket{\phi'_{r,y}}
    - \alpha_{r,y} \ket{\phi_{r,y}}\|^2 \\
\phantom{= \big\{} + \|\alpha'_{r,\overline{y}} \ket{\phi'_{r,\overline{y}}}
    - \alpha_{r,\overline{y}} \ket{\phi_{r,\overline{y}}}\|^2\big\}^{1/2}\\
\geq \big\{ (|\alpha'_{r,y}|^2 + |\alpha_{r,y}|^2 
    - 2 |\alpha'_{r,y}| |\alpha_{r,y}|)\\
\phantom{\geq \big\{} + (|\alpha'_{r,\overline{y}}|^2  
    + |\alpha_{r,\overline{y}}|^2 
    - 2 |\alpha'_{r,\overline{y}}|
|\alpha_{r,\overline{y}}|)\big\}^{1/2}\\
=  \big\{2 - 2(|\alpha'_{r,y}| |\alpha_{r,y}| 
    + |\alpha'_{r,\overline{y}}|
|\alpha_{r,\overline{y}}|)\big\}^{1/2}\\
 = \sqrt 2 \big\{1 - |\alpha'_{r,y}| |\alpha_{r,y}| 
    - |\alpha'_{r,\overline{y}}| |\alpha_{r,\overline{y}}|\big\}^{1/2}\\
\geq \sqrt 2 \big\{1 - |\alpha_{r,y}| 
    - |\alpha'_{r,\overline{y}}|\big\}^{1/2},
\end{array} \]
where the two inequalities follow from proposition~\ref{prop:1} 
and the fact that the absolute value of any scalar is at most one.

\noindent
Proof of (3):
By (2), (1), and 
proposition~\ref{prop:2},
\begin{eqnarray*}
1 - |\alpha_{r,y}| - |\alpha'_{r,\overline{y}}| 
  &\leq& {1 \over 2} \| \ket{\phi'_r} - \ket{\phi_r} \|^2 \\
  &\leq& 2 \left(\sum_{i=0}^{r-1} |\alpha_{i,y}|\right)^2
  \leq 2r \sum_{i=0}^{r-1} |\alpha_{i,y}|^2.
\end{eqnarray*}
Thus,
\begin{eqnarray*}
\sum_{y \in S}\left(1 - |\alpha_{r,y}| - |\alpha'_{r,\overline{y}}|\right)
&\leq& \sum_{y \in S} \left(2r \sum_{i=0}^{r-1} |\alpha_{i,y}|^2\right)\\
&=& 2r \sum_{i=0}^{r-1} \left(\sum_{y \in S} |\alpha_{i,y}|^2\right) 
= 2 r^2.
\end{eqnarray*}

Since,
\begin{eqnarray*}
&&\sum_{y \in S}\left(1 - |\alpha_{r,y}| - |\alpha'_{r,\overline{y}}|\right)
=N-\sum_{y \in S} |\alpha_{r,y}| - \sum_{y \in S}|\alpha'_{r,\overline{y}}| \\
&& \quad \geq N - \sqrt N \left(\sum_{y \in S} |\alpha_{r,y}|^2\right)^{1/2} 
       -  \sum_{y \in S}|\alpha'_{r,\overline{y}}| \\
&& \quad = N - \sqrt N -  \sum_{y \in S}|\alpha'_{r,\overline{y}}|,
\end{eqnarray*}
we have
\[N - \sqrt N -  \sum_{y \in S}|\alpha'_{r,\overline{y}}|
\leq \sum_{y \in S} (1 - |\alpha_{r,y}| - |\alpha'_{r,\overline{y}}|) 
\leq 2 r^2,\]
and (3) follows. \qed

\begin{theorem}
\label{thm:lower:unique}
Let $S$ be any set of $N$ strings, and $M$ be any oracle quantum
machine with bounded error probability.
Let $y \in_R S$ be a randomly and uniformly chosen element from
$S$. Put $O$ to be the oracle where $O(x)=1 \mbox{ if and only if }
x=y$.
Then the expected number of times $M$ must query $O$ in order to
determine $y$ with probability at least $1/2$ is at least 
\mbox{$\lfloor (\sin(\pi/8))\sqrt{N} \rfloor$}.
\end{theorem}

\proof
Let $S$ be any set of $N$ strings and $\cal C$ be any
configurationspace.  Let $\ket{\psi_0}$ be any superposition of
configurations, and $M$ any bounded-error oracle quantum machine. Given any
oracle~$O^\star$, assume that we run $M^{O^\star}$ for $s$ steps, and
assume that $M$ queries its oracle $O^\star$ $r$~times during the
computation.  Since we will only run $M$ using oracle $O^\star$ with
$O^\star (x)=0 \mbox{ if } x \notin S$, without loss of generality,
assume that $M$ never queries $O^\star$ on strings not in $S$.

First, consider the case that we run $M$ using the trivial oracle:
let $O$ be the oracle where $O(x)=0$ for all $x \in S$, and let
\begin{equation}
\eqlabel{eq:oracle-comp}
\ket{\psi_s} = A_s \dots A_1 \ket{\psi_0}
\end{equation}
be the unitary transformation corresponding to the computation of 
$M$ using oracle~$O$.

For all $i=1,\dots,r$, set $q_i$ to be the timestamp for $M$'s $i$'th
query, and set $q_{r+1}=s+1$.
Then \eq{eq:oracle-comp} can also be written as 
\begin{equation}
\ket{\phi_r} = U_r \dots U_1 \ket{\phi_0}
\end{equation}
where
$\ket{\phi_0}=A_{q_1-1} \dots A_1 \ket{\psi_0}$, 
and for all $i=1,\dots,r,$
$U_i=A_{q_{i+1}-1} \dots A_{q_i}$  and 
$\ket{\phi_i} = U_i \ket{\phi_{i-1}}.$
At the $i$'th query some configurations will query $O$, some will not.
For all $i=0,\dots,r-1,$ set $f_i(\ket{c})=x$ if $\ket{c}$ 
queries $x$ at the $i+1$'th query. 

Now, consider what happens if we flip one bit of the oracle bits: 
Given any $y \in S$, let $O'$ be the oracle where $O'(x)=1$ 
if and only if $x=y$. Then the computation of $M^{O'}$ corresponds to
the unitary transformation
$$\ket{\phi'_r} = U'_r \dots U'_1 \ket{\phi_0}$$
where $U'_i \ket{c}=U_i \ket{c}$ if $f_{i-1}(\ket{c}) \neq y$.

At the end of the computation of $M^{O'}$, we measure the 
superposition $\ket{\phi'_r}$ in order to determine the unknown~$y$. 
For each configuration $\ket{c} \in {\cal C}$, set 
$f_r(\ket{c})=x$ if, by measuring $\ket{c}$, $M$ answers that $x$
is the unknown $y$.

Set $\ket{\phi'_r} = \alpha'_{r,y}\ket{\phi'_{r,y}} +
\alpha'_{r,\overline{y}}\ket{\phi'_{r,\overline{y}}}$ where $\ket{\phi'_{r,y}}$
$(\ket{\phi'_{r,\overline{y}}})$ is the normalized superposition of
configurations where $f_r$ (does not) equals $y$.
Then $|\alpha'_{r,y}|^2$ is the probability that $M^{O'}$ correctly 
determines $y$. 
Since, by assumption, this probability is at least $1/2$, 
$$|\alpha'_{r,\overline{y}}| \leq {1 \over \sqrt 2}
\quad \mbox{ for all } \quad y \in S.$$
Furthermore, by Lemma~\ref{lm:main:lower}, 
\[N - \sqrt N - \sum_{y \in S} |\alpha'_{r,\overline{y}}| \leq 2 r^2.\]
Hence,
\begin{eqnarray*}
2 r^2 &\geq& N - \sqrt N - \sum_{y \in S} |\alpha'_{r,\overline{y}}| \\
   &\geq& N - \sqrt N - {1 \over \sqrt 2}N  \\
   &=& (1 - {1 \over \sqrt 2})N - \sqrt N,
\end{eqnarray*}
so 
\begin{eqnarray*}
r &\geq& \left\{(2-\sqrt 2) {N \over 4}
     - {\sqrt N \over 2}\right\}^{1/2}  \\
 &=& \left\{2 - \sqrt 2 - {2 \over \sqrt N}\right\}^{1/2} 
   {\sqrt N \over 2}\\
 &>& \left\{\sqrt{2-\sqrt 2} - {2 \over \sqrt N}\right\} 
   {\sqrt N \over 2} \\
 &=& \frac{\sqrt{2-\sqrt 2}}{2} \sqrt{N} -1 \\
 &=& (\sin(\pi/8)) \sqrt{N} -1 ,
\end{eqnarray*}
which proves the theorem.
\qed

Theorem~\ref{thm:lower:unique} gives a lower bound for finding a unique
feasible $y\in S$ using a bounded-error quantum machine. However, in most
applications we would expect that there will be more than one feasible
$y$, say $t$ such $y$'s. Furthermore, we might even not know if there is
a feasible $y$ or not. For the case $t\geq 1$, we have:

\begin{theorem}
Let $S$ be any set of $N$ strings, and $M$ be any bounded-error oracle quantum
machine.  Let~$A \subset_R S$ be a randomly and uniformly chosen subset 
of~$S$ of size~$t$, $t \geq 1$. 
Put $O$ to be the oracle where $O(x)=1$ if and only if 
$x \in A$.
Then the expected number of times $M$ must query $O$ in order to
determine some member $y \in A$
with probability at least $1/2$ is at least 
\mbox{$\lfloor (\sin(\pi/8)) \sqrt{\lfloor N/t\rfloor} \rfloor$}.
\end{theorem}

The proof of this theorem is almost identical to the proof of 
Lemma~\ref{lm:main:lower} and Theorem~\ref{thm:lower:unique}.
In Lemma~\ref{lm:main:lower}, equations (1) and 
(2) now hold for all subsets of $t$ strings. 
Hence, by choosing a largest number of such disjoint subsets from $S$,
say $T=\{X_1,\dots,X_{N_t}\}$ where $N_t = \lfloor N/t\rfloor$,
in the proof of (3), we obtain
$$N_t -\sqrt{N_t} - \sum_{X_i \in T} |\alpha'_{r,\overline{X_i}}| \leq 2r^2.$$
The remaining part of the proof is the same as the proof of
Theorem~\ref{thm:lower:unique}, only with obvious and minor changes. 

\section*{Acknowledgements}
We are grateful to Umesh and Vijay Vazirani for discussions
concerning classical approximate counting.
The~third author would like to thank Edmund Christiansen for
helpful discussions concerning recursion equations,
and Joan Boyar for helpful discussions in general.

\end{document}